\documentclass[useAMS,usenatbib]{mn2e}
\usepackage{amssymb,amsmath,epsfig,times,natbib,color}

\long\def\symbolfootnote[#1]#2{\begingroup%
\def\thefootnote{\fnsymbol{footnote}}\footnote[#1]{#2}\endgroup} 

\newcommand{\nustar}{\emph{NuSTAR}}

\title[Absorption and Intrinsic Variability in NGC~1365]{PCA of PCA: Principal Component Analysis of Partial Covering Absorption in NGC~1365}

\author[M. L. Parker et al.]{M. L. Parker,$^1$\thanks{Email: mlparker@ast.cam.ac.uk}
  D. J. Walton,$^{2}$
  A. C. Fabian$^{1}$
  and G. Risaliti$^{3,4}$\\
  $^{1}$Institute of Astronomy, Madingley Road, Cambridge, CB3 0HA\\
  $^{2}$California Institute of Technology, 1200 East California Boulevard, Pasadena, CA 91125, USA\\
  $^{3}$INAF--Osservatorio Astrofisico di Arcetri, Largo Enrico Fermi 5, 50125 Firenze, Italy\\
  $^{4}$Harvard-Smithsonian Center for Astrophysics, 60 Garden St, Cambridge, MA 02138, USA\\
 }
\date{}

\begin{document}

\maketitle

\begin{abstract}
We analyse 400 ks of \emph{XMM-Newton} data on the active galactic nucleus NGC~1365 using principal component analysis (PCA) to identify model independent spectral components. We find two significant components and demonstrate that they are qualitatively different from those found in MCG--6-30-15 using the same method. As the variability in NGC~1365 is known to be due to changes in the parameters of a partial covering  neutral absorber, this shows that the same mechanism cannot be the driver of variability in MCG--6-30-15. By examining intervals where the spectrum shows relatively low absorption we separate the effects of intrinsic source variability, including signatures of relativistic reflection, from variations in the intervening absorption.
We simulate the principal components produced by different physical variations, and show that PCA provides a clear distinction between absorption and reflection as the drivers of variability in AGN spectra. The simulations are shown to reproduce the PCA spectra of both NGC~1365 and MCG--6-30-15, and further demonstrate that the dominant cause of spectral variability in these two sources requires a qualitatively different mechanism.
\end{abstract}


\section{Introduction}

Principal component analysis (PCA) is a sophisticated tool for finding patterns in complex datasets. By applying PCA to a set of spectra from the same source but different times, model independent components which describe the source variability as efficiently as possible can be found \citep{Mittaz90}. The main advantage of this method over calculating the RMS spectrum \citep{Taylor03} comes when more than one spectral component is varying, for example when the continuum flux is modified by variable absorption, or when a strong relativistic reflection component is present. While other methods can return the total variability as a function of energy, PCA will return the variability in each separate varying component.

\citet{Miller08} use PCA to argue that the spectral shape and variability in the AGN MCG--6-30-15 can be explained using a partial covering absorption model, where the source is obscured by intervening material which varies in column density and ionisation. However, a more detailed analysis using a similar method but with a larger dataset by \citet{Parker13} found that the variability was most likely dominated by variations intrinsic to the source, in the form of a power law continuum and a separate relativistically blurred reflection component. In general, it can be difficult to distinguish between relativistic reflection and absorption models in AGN from spectral fitting alone (see e.g. the comparison of the two in MCG--6-30-15 by Marinucci et al., submitted), as these can often give statistically equivalent fits. To break this degeneracy we must consider the spectral variability, as the two models make different predictions for how the spectrum should change. 

In this work we analyse the variability of a source which unambiguously shows evidence of strong variable absorption, NGC~1365. Recent analysis of the joint \nustar{} \citep{Harrison13} and \emph{XMM-Newton} \citep{Jansen01} spectra of NGC~1365 by \citet[herafter R13]{Risaliti13} and Walton et al., (submitted, hereafter W14) found that the relativistic reflection features of the spectrum in this source were independent of the level of absorption, and that models without a blurred reflection component were disfavoured both physically and statistically. However, the intrinsic spectrum of the source appears to be relatively stable, with the spectral variability being dominated by absorption processes. This combination of strong, variable absorption, a relatively constant underlying spectrum and an abundance of data makes NGC~1365 an ideal source with which to study absorption variability.

\section{Observations, Data Reduction and Analysis}

NGC~1365 was observed four times simultaneously by \emph{XMM-Newton} and \nustar{}, in 2012 July and December  and 2013 January and February. In this work, we use only the higher count rate \emph{EPIC-pn} data. We use $\sim40''$ circular extraction regions for all spectra, and the data are filtered for background flares. For full details of the data reduction used, see W14. 
We divide the observations into spectra from 10~ks intervals, as this gives a good compromise between signal and timing resolution. We note that the results described in Section~3 do not differ significantly if we use the same intervals as W14, where they were selected manually based on flux and hardness. Unlike W14, we do not use an annular region for the single period where the flux is slightly above the recommended limit for avoiding pile-up.
The reliability and power of PCA is strongly dependent on the count rate, and pile-up should not have any noticeable effect on the results of such an analysis as we are not attempting to make extremely detailed and sensitive measurements of spectral features, such as the iron line profile. To confirm this, we repeated the analysis excluding this interval, and found no significant differences. Excising the core of the PSF in W14 was a deliberately conservative reduction, as the relative grade pattern distribution suggested pile-up effects were actually negligible across the standard XMM bandpass (0.3-10keV). The MOS detectors were the ones primarily effected by pile-up. 

The spectra were processed and analysed as described in \citet{Parker13}. Normalised, residual spectra of deviations from the mean are calculated, and binned logarithmically into 70 energy bins. Singular value decomposition (SVD) is then used to calculate the principal components. The resulting spectra show the correlations between different energy bins, so a flat line would imply that all bins are equally variable and strongly correlated, a suppression towards zero implies shows that the component is less variable at that energy, and a bins below zero show an anticorrelation to those above. This is complicated by the requirement of the analysis that the components are orthogonal, i.e. their dot products must equal zero, which means that higher order components cannot be entirely positive. These components must instead be regarded as corrections to the lower order components, but while this makes the interpretation slightly more difficult it does not compromise the analysis results.

All errors quoted or shown are at one standard deviation unless otherwise stated. Errors on the principal component spectra are calculated by perturbing the input spectra, as discussed in \citet{Miller07}.

\section{Principal Component Analysis Results}

\begin{figure}
\centering
\includegraphics[width=\linewidth]{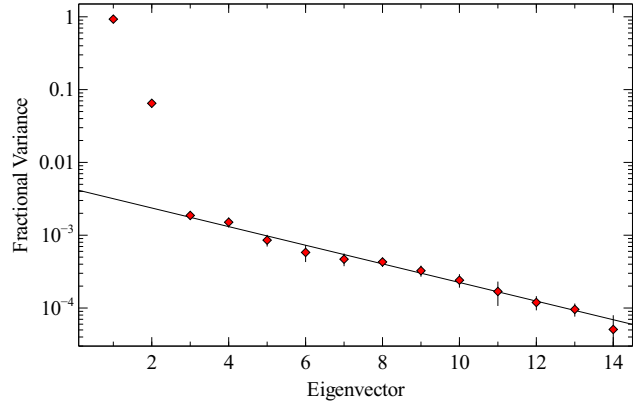}
\caption{LEV diagram, showing the fractional variability of each eigenvector. Error bars are plotted, but are smaller than the first two points. The black line shows the best-fit geometric progression, fit to the points from 3 onwards. This shows the expected contribution of noise. The first two components are highly significant ($>50$ standard deviations from the expected noise level).}
\label{lev}
\end{figure}

We first applied our PCA method to the full set of XMM observations obtained in late 2012/early 2013.
Fig.~\ref{lev} shows the log-eigenvalue (LEV) diagram for this analysis. This shows the fractional variability in each of the principal component produced, and can be used to determine the statistical significance of the components returned by the analysis \citep{Jolliffe02}. Components due to noise are predicted to decay geometrically, and we find that only the first two components deviate from this prediction significantly. As a simple test of significance, we calculate the Pearson correlation coefficient between adjacent bins for each component and find that the probability of either component arising by noise (in which case, adjacent bins would not be correlated) is less than $10^{-30}$.

\begin{figure*}
\centering
\includegraphics[width=12cm]{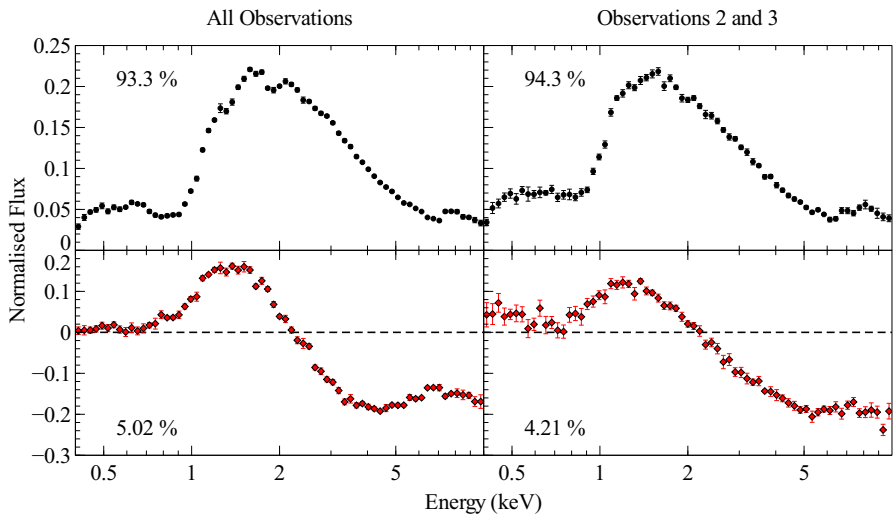}
\caption{Left: Spectra of the two significant principal components of NGC~1365, and fractional variability in each component. Both are suppressed at low energies ($E<1$~keV) by the presence of diffuse thermal emission. The first component (black, top) is responsible for $\sim93$~per cent of the variability, and the second (red, bottom) for $\sim6$ per cent. Error bars are plotted for both spectra, but are only visible for component 2. Right: as left, but for the second and third observations. The largest differences occur in the shape, size and energy of the high energy feature, which is broader, stronger, and lower in energy than in the full analysis.}
\label{pcaspecs}
\end{figure*}

\begin{figure}
\centering
\includegraphics[width=\linewidth]{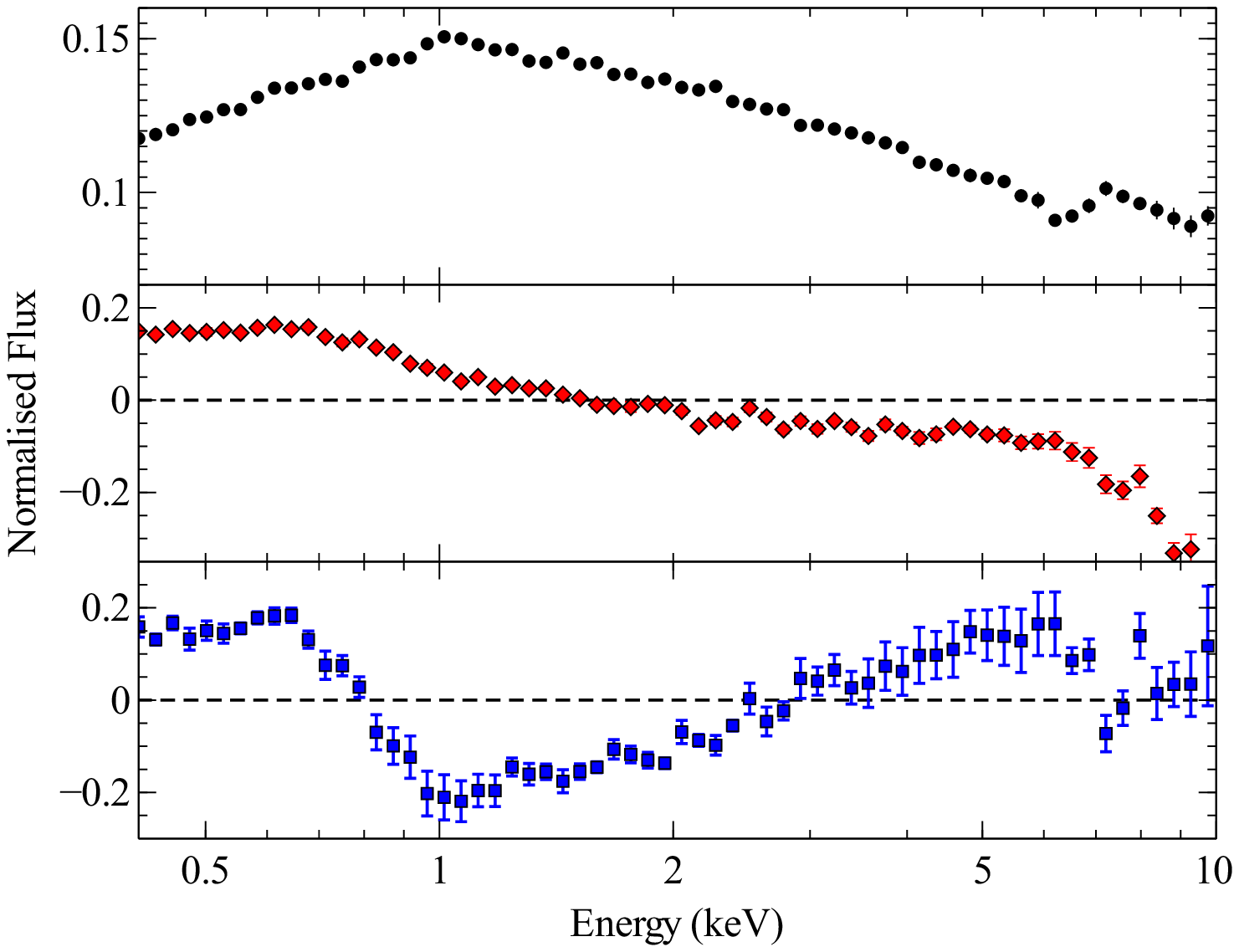}
\caption{Spectra of the first three PCs reproduced from the analysis of MCG--6-30-15 presented in \citet{Parker13}. Errors are plotted for all components, but are smaller than the majority of the points in the first two spectra.}
\label{mcg6pcaspecs}
\end{figure}

The spectra of the two significant principal components (PCs) are shown in  Fig.~\ref{pcaspecs}. Comparing these components with those found for MCG--6-30-15 (shown in Fig.~\ref{mcg6pcaspecs}) reveals significant differences, which we discuss in Section~\ref{simulations}.  Both PCs are suppressed at energies below $\sim1.5$keV, as expected, due to the presence of diffuse thermal emission around the nucleus, which has been resolved using \emph{Chandra} \citep{Wang09}. This emission almost entirely dominates the spectrum below $\sim$2~keV. Previous analysis of the spectral variability in this source has largely been restricted to the 2.5-10~keV energy band, to exclude the effects of dilution from the surrounding material. We analyse the data again over this energy band, with the same number of energy bins, and find the same two components with no significant differences. For the remainder of this work we consider the full energy range.

The first principal component (PC1) from NGC~1365 is dominated by an excess between 1--5~keV. This is the energy range over which the effects of absorption are prominent, and a strong absorption edge is visible at 7~keV, but no other significant features are visible. It therefore seems likely that this component corresponds to the large changes in the column density of the absorber seen in this source.
The second PC shows an anticorrelation between low and high energies, as found in the second component of MCG--6-30-15, where it was found to be well explained by pivoting of the power law. However, this component is qualitatively different from that found in MCG--6-30-15, again showing the absorption edge at 7~keV, and without the steepening at high energies. We conclude that this component is also caused by a change in the properties of the absorber, and discuss this further in section~\ref{simulations}.

The second and third observations from the joint \emph{NuSTAR} and \emph{XMM-Newton} campaign (XMM obs. IDs 0692840301 and 0692840401) revealed the source in an almost completely uncovered state, which had not previously been seen. W14 found that during these observations the column density dropped to values as low as $1.3\pm 0.3 \times 10^{22}$cm$^{-2}$, with an average of $\sim 7\times 10^{22}$cm$^{-2}$. 
By examining these two observations in isolation we hope to examine the variability of the source itself, rather than the intervening material. In general, intrinsic variability will be distinguishable when it is strong relative to the variability in absorption. For this, we require observations where the absorption is constant, not absent, although a higher signal to noise will result from using the higher flux states.


\begin{figure}
\centering
\includegraphics[width=\linewidth]{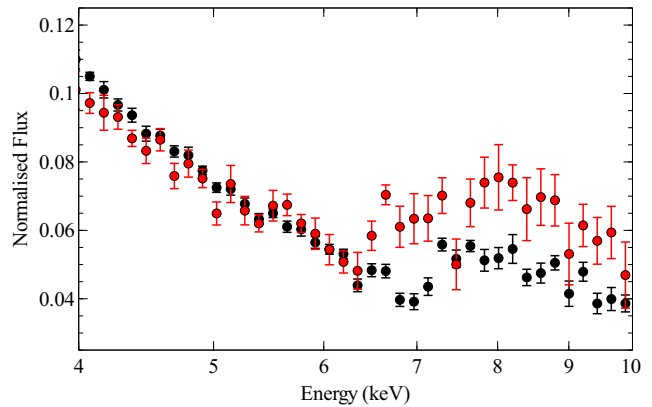}
\caption{The same components shown in the top panels of Fig.~\ref{pcaspecs}, but focussing on the high energy (4--10~keV) spectra. This shows the much broader, lower energy feature found in the restricted data set (red) compared with the smaller, sharper 7~keV feature found when we include all the available data (black).}
\label{int3comparizoom}
\end{figure}

As in the analysis of the full dataset we find two significant components, although the second is responsible for a considerably smaller fraction of the variability in this case. Only the first component, shown in the middle panel of Fig.~\ref{pcaspecs}, displays significant differences from that found in the full analysis. Although there are some differences at lower energies, the major change comes in the shape and energy of the high energy feature. In the analysis of the full dataset, this was a sharp absorption edge at 7~keV. However, in the restricted analysis this feature is broader, deeper, and has a minimum at a lower energy ($\sim6$~keV), as shown in Fig.~\ref{int3comparizoom}. If this was being produced by absorption as in the full analysis we would expect the 7~keV feature to get weaker as the column density dropped, not stronger, so this suggests that the feature is intrinsic to the source spectrum in this case. We conclude that this feature is likely to be produced by variations in the power law continuum flux which are suppressed by a relatively constant relativistic reflection component.

\section{Simulations}
\label{simulations}

We adopt a different approach from \citet{Parker13} to determine the nature of the principal components. \citeauthor{Parker13} fit models to extremal spectra, and compared the normalisations of the PCs with the results of spectral fitting to the same set of spectra. Instead, we create simulated spectra using simplified models of AGN spectra, allowing certain parameters to vary. These spectra are then analysed in the same way as the real spectra, and the resultant components compared with those found from observations.

\subsection{Method}
We use simple spectral models in \textsc{xspec} version 12.8.1, based on the best fit models used by W14 (for NGC~1365) and Marinucci et al. (for MCG--6-30-15), to create simulated 10~ks \emph{XMM-Newton} EPIC-pn spectra. We perturb the parameters of interest (i.e. those we suspect are driving the variability) randomly between the limits discussed in the respective papers, and generate a fake 10~ks spectrum for each set of parameters using the \textsc{xspec} `fakeit' command. These spectra are then analysed using the same PCA code as the real data.

\subsection{Absorption modelling}
\label{absmodelling}
The variability described by W14 for NGC~1365 is dominated by changes in the column density of the partial covering absorber, which varies from $(1.3\pm 0.3) \times 10^{22}$ cm$^{-2}$ to $(26.4 \pm 0.9) \times 10^{22}$ cm$^{-2}$. The covering fraction of the absorber also varies, but by a smaller amount, between $0.80\pm 0.04$ and $0.972\pm0.002$. We use a simplified version of the model discussed by W14, neglecting both reflection and ionised absorption. Because the PCA is calculated based on fractional residuals for each spectrum, constant absorption components have no effect on resultant spectra, and can be neglected. In general, constant additive components such as relativistic reflection will also contribute to the PCA components identified, but in the case where the spectral variability is strongly dominated by the multiplicative absorption component, which applies equally to both relativistic reflection and the power law continuum, they can safely be ignored.

Therefore, to simulate the effects of this variability we use the \textsc{xspec} model \textsc{bbody+pcfabs*powerlaw}. The black body component is to mimic the effects of the diffuse emission suppressing the source variability at low energies. We fix the temperature of this component to 0.1~keV, and keep it constant. The power law is also kept constant, and the photon index is fixed at 2.
\begin{figure}
\centering
\includegraphics[width=\linewidth]{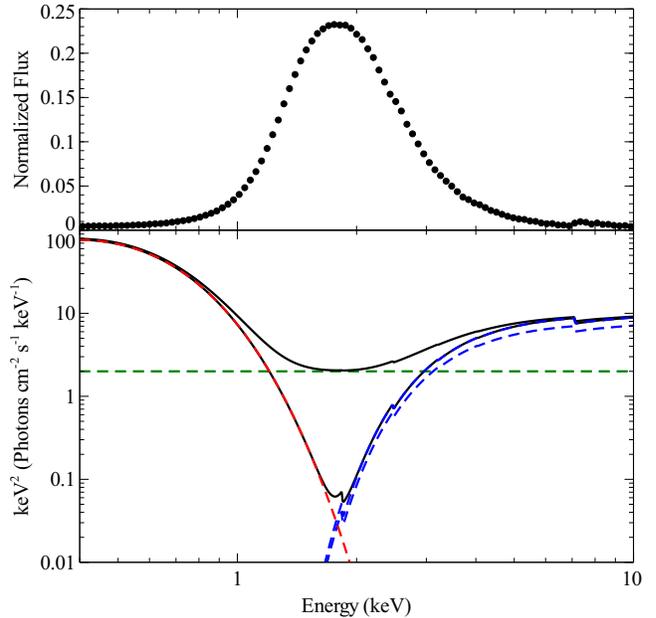}
\caption{Top: Spectrum of the one significant component returned from a simulation where the covering fraction of a variable absorber is allowed to change. There is almost no variability at high energies, and at low energies the spectrum is dominated by the black body component, drowning out the source variability. Bottom: Model spectra used to simulate the principal components of NGC~1365. The two black solid lines show the total models, for covering fractions of 0.8 and 1 (upper and lower, respectively). The blue dashed lines on the right show the absorbed component, and the green dashed line shows the unabsorbed component, for the case where the source is not fully covered. The red dashed line shows the constant black body component.}
\label{cvrfracboth}
\end{figure}
Initially, we allow only the covering fraction of the absorber to vary (between 0.8 and 1, although the range only affects the amplitude of the signal, not the spectral shape in this case), keeping the column density constant at $10^{23}$~cm$^{-2}$. This produces a single significant PC, which is shown in the upper panel of Fig.~\ref{cvrfracboth}. While this component is generally similar to the first PC found from the NGC~1365 data, it only shows a weak absorption edge at 7~keV, which is clearly visible in the data (Fig.~\ref{pcaspecs}). The lower panel of Fig.~\ref{cvrfracboth} shows the two extreme models for the minimum and maximum covering fractions, from which it can clearly be seen that the variability caused by changes in the covering fraction is concentrated at intermediate energies, as demonstrated by the PCA spectrum.

\begin{figure}
\centering
\includegraphics[width=\linewidth]{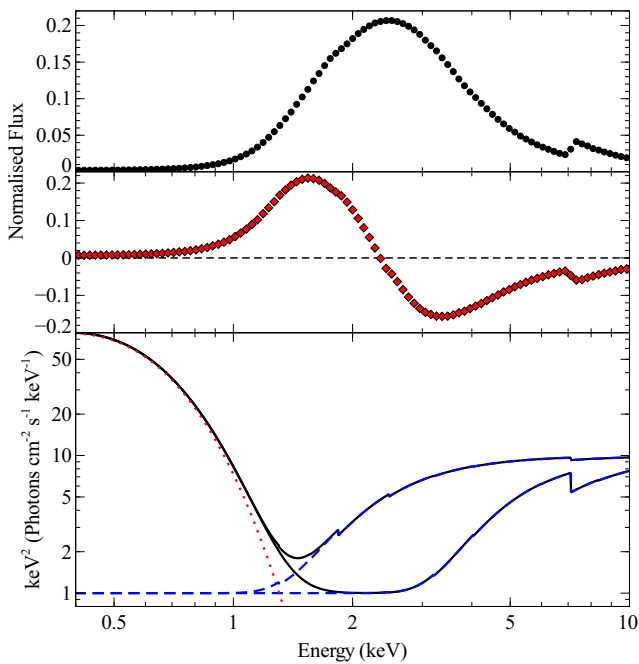}
\caption{Top and Middle: Spectra of the two PCs found from a simulation where both the covering fraction and column density of a partial covering absorber are allowed to vary, with the column density causing most of the spectral variability (see text). Bottom: Model spectra used to simulate the effects of varying column density. The black solid lines show the minimum and maximum spectra, the red dotted line shows the black body component, and the blue dashed lines show the partially covered absorbed power law.}
\label{1365sim}
\end{figure}

When we include the stronger variations in the column density (from 5--25$\times10^{22}$~cm$^{-2}$) a second component is returned, as shown in Fig.~\ref{1365sim}, and the shape of the primary component changes. These two components are an excellent qualitative match for those found from the data for NGC~1365. The sharp 7~keV feature found in the PCs from NGC~1365 is also found in the simulated spectra, as is the difference in peak energy between the two components.
Interestingly, both of these components are produced even if only the column density is allowed to vary. The need for the second component when the column density changes is due to the changes this causes in the spectral shape of the absorbed component.

The bottom panel of Fig.~\ref{1365sim} shows the model components when only the column density has been allowed to vary, over the same range as previously, with a covering fraction of 0.9. Unlike the simple case where only the covering fraction changes, the energy dependence of the variability here changes depending on the column density. This therefore means that changes in the column density cannot be described adequately by a single principal component, necessitating a `correction factor', in the form of the second component. We note that a combination of these two components can also be used to make the component shown in Fig.~\ref{cvrfracboth}, where only the covering fraction varies. This means that the relatively weak variations in the covering fraction can be adequately described without the introduction of a third principal component. Because of this, the components returned from our simulations do not differ significantly in either spectral shape or fractional variability when we include or exclude variations in the covering fraction. This should not hold in general as the low energy differences between the two cases should be significant, however these are obscured by the thermal component in this model.

\begin{figure}
\centering
\includegraphics[width=\linewidth]{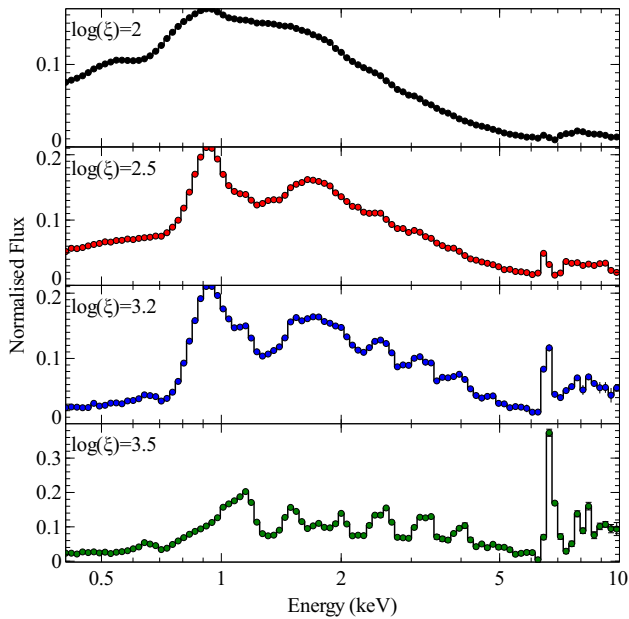}
\caption{Principal components returned from simulations of variable partial-covering ionised absorption, where the covering fraction is kept constant and the column density is allowed to change. The four panels correspond to four different ionisation states, shown in the top left corners.}
\label{ionisedabssim}
\end{figure}

Finally, we consider the effect of variable ionised absorption. For this, we use the \textsc{zxipcf} model of \citet{Reeves08}. We fix the covering fraction at 0.5 and vary the column density between (0.5--5)$\times10^{22}$, for a range of ionisation parameters. The results of this are plotted in Fig.~\ref{ionisedabssim}, and show a strong dependence of the spectral shape on the ionisation of the absorber. Unlike the neutral case, varying the covering fraction instead of the column density does not produce a significant difference in the shape or number of components, so we only show the results for column density. None of the PCs returned from the data for either source appear to be a good match to the highly ionised absorption components, and we conclude that while there is evidence of ionised absorption in both sources \citep[e.g.][]{Otani96,Risaliti05} it is not the main driver of spectral variability in either object.

\subsection{Reflection modelling}
\label{refmodelling}

\begin{figure}
\centering
\includegraphics[width=\linewidth]{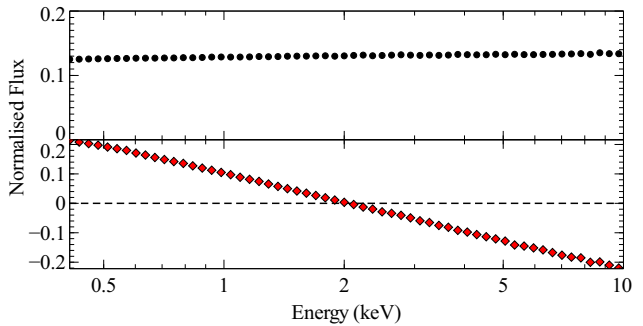}
\caption{Spectra of the first (top, black) and second (bottom, red) PCs found from a simulation of a power law, where the normalisation and photon index are both allowed to vary. The first component corresponds to changes in the normalisation, and the second to changes in the photon index.}
\label{powerlawsim}
\end{figure}

To investigate the effects of intrinsic source variability, we start with a simple powerlaw model. The normalisation is allowed to vary by a factor of 5, and the photon index is allowed to vary between 1.9 and 2.1. This returns two significant PCs, shown in Fig.~\ref{powerlawsim}. As expected for such a simple model, the two components returned are straight lines. The first component is flat and positive, and corresponds to changes in the normalisation of the powerlaw. The second PC shows an anticorrelation between low and high energies, and corresponds to changes in the photon index of the powerlaw.

\begin{figure}
\centering
\includegraphics[width=\linewidth]{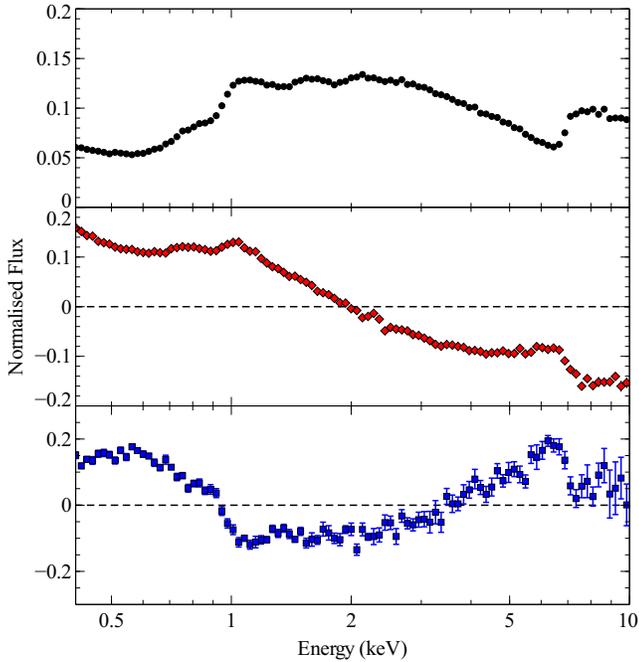}
\caption{Spectra of the three PCs found from a simulation where a primary power law is allowed to vary in normalisation and photon index, along with a relativistic reflection component which is allowed to vary only in normalisation and more weakly than the power law. The reflection fraction is exaggerated here to make the features in component one stronger, and we neglect the effect of distant reflection.}
\label{mcg6sim}
\end{figure}

If we introduce a constant blurred reflection component and allow the power law to vary as before, the two components returned show significant differences (see the upper two panels of Fig.~\ref{mcg6sim}, note that the component shown in the figure are for the case where the reflection component varies by a factor of 2 (see below), but we do not find any significant differences in the spectral shape between the two cases). We model the reflection component using the \textsc{xillver} reflection model \citep{Garcia13}, convolved with the \textsc{kdblur} relativistic blurring model. We fix the emissivity index at 3, the inner and outer radii at 1.235 and 400~$R_\textrm{G}$, respectively, the inclination at 40 degrees, the photon index at 2, the ionization parameter at 10~erg~cm~s$^{-1}$, the iron abundance at one, and set the normalisation so that the total flux in the reflection component is roughly equal to the mean flux of the power law, over the 0.4--10~keV band.
Both the components produced are suppressed towards zero at the energies of the iron line and soft excess. In the first component (responsible for 98.6 per cent of the variability) this results in a flat line with a reflection spectrum subtracted, and in the second component (responsible for 1.4 per cent) it causes two breaks to appear in the spectrum, one at the energy of the iron line and one at the start of the soft excess. All of these features are simply due to the relatively constant reflection component suppressing the variability in the powerlaw. We note that all of these features are visible in the first two components found for MCG--6-30-15 (Fig.~\ref{mcg6pcaspecs}), and that the iron line features are very distinct from the sharp absorption edges seen in the PCA spectra of NGC~1365.

We then consider the effect of allowing the reflection component to vary. We allow the reflection normalisation to vary within a factor of two, compared to the factor of five allowed for the power law, consistent with observational measurements \citep[e.g.][]{Fabian02} and theoretical predictions from light-bending models \citep[e.g.][]{Martocchia96}, where the continuum flux varies considerably more than the reflected emission. This results in a third component being produced (shown in the bottom panel of Fig.~\ref{mcg6sim}), which shows strong correlated iron line and soft excess features. This PC is a good qualitative match for the third component found in MCG--6-30-15, shown in the bottom panel of Fig.~\ref{mcg6pcaspecs}, which shows the same key features of relativistic reflection. Although the spectral shape of the first two components does not change significantly when the reflection component is allowed to vary, the fractional variability in them does. As expected, the variability in the primary component drops to 96.9 per cent, but the variability in the second component actually increases, to 2.5 per cent. The variability in the third component is then 0.5 per cent. This increased variability in the second component suggests that a large part of the reflection variability is actually being attributed to this component, rather than the third.

\begin{figure}
\centering
\includegraphics[width=\linewidth]{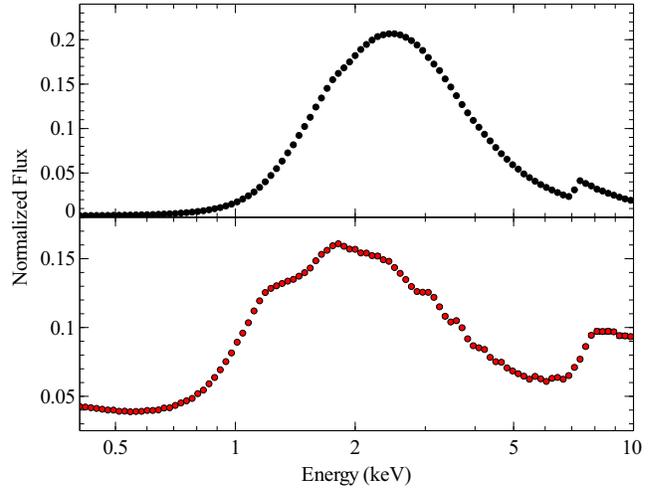}
\caption{Top: first component returned from a simulation of partial covering absorption, where changes in the column density dominate the spectral variability (reproduced from Fig.~\ref{1365sim}). Bottom: first (and only) component returned when the absorption variability is assumed to be negligible, and the power law varies in the presence of a constant relativistic reflection component, using the parameters found by R13 and W14.}
\label{1365refsim}
\end{figure}

To try and reproduce the PCs found from the relatively unabsorbed observations of NGC~1365 (Fig.~\ref{pcaspecs}, right panel) we modify the reflection model to be closer to that found by spectral modelling by R13  and W14. The inclination is changed to 60 degrees, the emissivity index to 7, the ionization parameter to 50~erg~cm~s$^{-1}$, and the iron abundance to 5. The absorption is assumed to be zero for this model, and everything apart from the power law normalisation is fixed. The power law flux is then allowed to vary by a factor of 2. We also include the black body component described in Section~\ref{absmodelling}. From this simulation we find a single significant component, which is shown in the lower panel of Fig.~\ref{1365refsim}. When compared to the upper panel, which shows the first PC from the absorption model, there are several significant differences. The high energy feature, rather than being caused by a sharp 7~keV absorption edge, is now caused by suppression of the source variability by the constant iron line which leads to a much broader, lower energy feature. The component peaks at a lower energy, and there is more variability at low energies as well. Finally, there is more structure in the spectrum, including two breaks between 1--2~keV. All of these features can also be seen in the PCs found from the data over the second and third observations, so we conclude that we are seeing a stronger relative contribution from variations intrinsic to the source.

\section{Discussion}

The differences between the predictions of simple reflection and absorption models for AGN variability are very striking when analysed using PCA. The components produced by absorption variability show a sharp absorption edge at 7~keV, which is absent in the reflection model, and the spectral shape of the returned components depends strongly on the column density. Reflection variability, on the other hand, shows a correlated soft excess and iron line, along with suppression of the primary power law component at the same energies if the reflection component does not follow a 1:1 relation with the continuum. 
The simulations described in section~\ref{simulations} show that simple models for the variability in NGC~1365 and MCG--6-30-15 can qualitatively reproduce the principal components found from observational data. The spectral features of the PCs produced by the absorption and reflection simulations match very well with the features of the PCs from observational data of NGC~1365 and MCG--6-30-15, respectively, and are clearly distinct from one another. This demonstrates that PCA can be a powerful method for distinguishing these two potential origins for spectral variability in AGN.

The model proposed by \citet{Miller13} for the spectral shape observed in NGC~1365 uses a combination of partial covering absorption and scattering to produce the relatively constant spectral component. While the PCA results do not rule out such a model for this source, it does constrain it. Notably, the absorption zone responsible for the broadening must be independent of the absorption causing the majority of the spectral variability, otherwise no broad suppression feature would be present in the primary component found during the low absorption intervals. Secondly, as in the reflection model, this component must be much less variable than the continuum on 10~ks time scales. While this is reasonable for the scattered flux in such a model, the absorbed component is also required to be constant which implies a strong anticorrelation between the continuum flux and the covering fraction of the absorber. The case of MCG--6-30-15 is simpler, as the soft variability is not obscured by thermal emission. In the model proposed by \citet{Miller08} the soft excess arises from a steep power law, modified by absorption, while the constant iron line feature arises from a separate component. Therefore, in this model there should be no correlation between the soft excess and iron line, and the continuum variability should not be suppressed by the soft excess at low energies. As we have demonstrated here and in \citet{Parker13} this is not the case and the reflection interpretation is clearly preferred.

We note that on short time scales relativistic reflection should be correlated with the continuum flux, with a short lag between the two (around $\sim100$~s for a $10^6 M_\odot$ black hole). These lags have been found in several sources \citep[including MCG--6-30-15 and NGC~1365,][Kara et al., in preparation]{Emman11}, and occur at frequencies on the order of a few $10^{-4}$Hz. The results of this analysis and that of \citep{Parker13} which show that the flux of the relativistic reflection component is at least partially independent of the continuum flux on time scales of $\sim10$~ks do not conflict with this, requiring only that the reflection component is less variable than the continuum and that not all of the variability is correlated. In fact, it is likely that PCA and lag analyses can be complementary, as by definition lag analyses find variability that is correlated with the continuum, while higher order PCA terms can reveal variability that is weakly correlated or uncorrelated with the continuum (assuming that the primary component corresponds to the continuum, as in MCG--6-30-15). It should be possible to use PCA to probe different time scales within the same source, to see if and how the components change with the amplitude and sign of the lag. Also, it has been shown that the frequency and amplitude of the lags scales with mass \citep{DeMarco13,Kara13}, so by probing the same time scales in different sources, we can also investigate these changes.

We also investigate the possibility of distinguishing absorption and intrinsic variability in the same source, by using PCA to investigate the NGC~1365 data during different intervals. As shown in the right panel of Fig.~\ref{pcaspecs}, the shape of the primary component changes significantly when we investigate the observations where the source is minimally obscured, showing a broad iron line feature not found in the analysis of the full data set. This feature is broader, lower energy, and stronger than that predicted for absorption variability, and becomes more prominent when the column density drops. Also, there appears to be no reduction in the high energy variability, as would be expected if this was due to absorption variations and the column drops. These arguments point strongly to intrinsic continuum variability, in the presence of a relatively constant reflection component. This is consistent with findings in other AGN \citep[e.g.][]{Miniutti03,Ponti06}, where the lack of reflection variability is attributed to light-bending effects. We note that in sources where the low energy variability is not diluted by diffuse thermal emission it should be easier to untangle the intrinsic variability with relativistic reflection and the absorption variability, as only intrinsic variability should be suppressed at low energies, due to the presence of a relatively constant soft excess, as seen in MCG--6-30-15. We have modelled the intrinsic variability slightly differently between the two sources, keeping the reflection component constant in NGC~1365 and allowing it to vary in MCG--6-30-15. This is for simplicity, as no separate reflection component is detected, rather than indicating that the variability necessarily is different, or that the reflection component in NGC~1365 is constant. It should be possible to put an upper limit on the strength of the reflection variability based on this non-detection, but this would require more complex and precise modelling that is beyond the scope of this work.

We note that, in the case of NGC~1365, the primary components returned from variable absorption and intrinsic continuum variations have a similar overall shape (Figs.~\ref{pcaspecs},~\ref{1365refsim}). This is largely due to the presence of the diffuse thermal emission in this source, which obscures the soft excess from the relativistic reflection component. Were the thermal emission not present, both models would show increased low energy variability, but only the intrinsic variability would show suppression caused by the soft excess. This has two implications: firstly, it should be easier in general to distinguish these causes of variability in other AGN (where the low energy spectrum can be accurately measured), and secondly, that we can only see the dominant mechanism, not both, in NGC~1365. Because the differences between the components produced by the two mechanisms is relatively small, the analysis will not produce separate PCs for each, rather, it attributes the variability of both physical components to the same PC. The PC produced will then appear most like whichever mechanism is driving the majority of the variability on that time scale, with a second, relatively weak component accounting for the differences between the two models. Unfortunately, this component is likely to be lost in the noise. In general, this should not be a problem in sources where the soft excess can be observed, so separate components should in theory be retrievable.

We have neglected the effects of distant reflection in our simulations, to keep the models as simple as possible. Distant reflection should, in general, be variable only on very long time scales, and usually accounts for only a small fraction of the flux so its effect on the results should be minimal. The largest effect would come at 6.4~keV, the energy of the iron line, where the variability of the other spectral components will be suppressed. However, in the PCA spectra presented here it appears that this effect is negligible compared to those of relativistic reflection and absorption.

\section{Conclusions}
We have examined the spectral variability of NGC~1365 using PCA to isolate two model independent components which explain $\sim99$ per cent of the variability, with the remainder attributable to noise. These components are qualitatively different from those found in a similar analysis of MCG--6-30-15, demonstrating that the main driver of variability in the two sources must have a different origin. While the variability in NGC 1365 is dominated primarily by variable absorption, when focusing specifically on the least obscured periods we also see evidence for an increased contribution from intrinsic variability.

Simulated spectra, based on simple models of AGN spectra, are used to investigate the origin of the variability in these two sources, and show that we can reproduce the PCA spectra from the data using absorption and reflection models for NGC~1365 and MCG--6-30-15, respectively. We also simulate the principal components produced by highly ionised absorption, which gives another unique prediction. We do not find evidence for variable ionised absorption in either source, although there is strong evidence for the presence of such absorption. This indicates that the ionised absorbers are only weakly variable, and not responsible for the large changes in spectral shape seen.

We show that PCA can be extremely useful in distinguishing between absorption and reflection variability in AGN, as while these two origins can produce similar spectra their variability properties are different, and  these can potentially be disentangled in a model independent way using this technique.

\section*{Acknowledgements}
The authors would like to thank the anonymous referee for their helpful comments. MLP acknowledges financial support from the Science and Technology Facilities Council (STFC). 
\bibliographystyle{mn2e}
\bibliography{ngc1365_paper}
\end{document}